\documentclass[showpacs,amsmath,amssymb,preprint,showkeys]{revtex4}
\usepackage{epsfig,dcolumn,bm}

\begin{document}

\title{A possible $\Omega\pi$ molecular state}
%\footnote{The project supported in part by National Natural Science
%Foundation of China under Grant No. 10475087}}

\author{W.L. Wang$^1$}
\author{F. Huang$^2$}
\author{Z.Y. Zhang$^3$}
\author{Y.W. Yu$^3$}
\author{F. Liu$^1$}
\affiliation
{\small $^1$Institute of Particle Physics, Central China Normal University, Wuhan 430079, China\\
$^2$CCAST (World Laboratory), P.O. Box 8730, Beijing 100080, China \\
$^3$Institute of High Energy Physics, P.O. Box 918-4, Beijing
100049, China}

\begin{abstract}
The structure of $\Omega\pi$ state with isospin $I=1$ and
spin-parity $J^p=3/2^-$ are dynamically studied in both the chiral
SU(3) quark model and the extended chiral SU(3) quark model by
solving a resonating group method (RGM) equation. The model
parameters are taken from our previous work, which gave a
satisfactory description of the energies of the baryon ground
states, the binding energy of the deuteron, the nucleon-nucleon
($NN$) scattering phase shifts, and the hyperon-nucleon ($YN$) cross
sections. The calculated results show that the $\Omega\pi$ state has
an attractive interaction, and in the extended chiral SU(3) quark
model such an attraction can make for an $\Omega\pi$ quasi-bound
state with the binding energy of about several MeV.
\end{abstract}

\pacs{12.39.-x, 21.45.+v, 11.30.Rd}

\keywords{$\Omega\pi$ quasi-bound state, quark model, chiral
symmetry}

\maketitle

\section{Introduction}

The possible new resonance state is always an interesting topic in
both the experimental and theoretical physics, and it can help us to
catch more details about the hadron-hadron and quark-quark
interaction. In 2001, Gao {\it et al.} followed the idea of Ref.
\cite{brodsky90} and estimated the QCD van der Waals attractive
force of the $N\phi$ system \cite{hygao01}. They claimed that the
QCD van der Waals attraction between $N$ and $\phi$, mediated by
multi-gluon exchanges, can be strong enough to form a bound $N\phi$
state with a binding energy of about $1.8$ MeV. At the same time
they pointed out that it is possible to search for such a bound
state using the $\phi$ meson below threshold quasi-free
photo-production kinematics experimentally. Using a simple model,
the authors calculated the rate for such subthreshold quasi-free
production process using a realistic Jefferson Laboratory luminosity
and a large acceptance detection system. They concluded that such an
experiment is feasible. Recently, we have performed a dynamical
study of the structures of $N\phi$ states with $J^{p}=3/2^-$ and
$J^p=1/2^-$ in the chiral quark model by solving a resonating group
method (RGM) equation \cite{fhuang06nphi}. Our results show that the
$N\phi$ states have attractive interaction, and in the extended
chiral SU(3) quark model such attraction, dominantly provided by the
quark and $\sigma$ field coupling, can make $N\phi$ states to be
bound with several MeV binding energies.

The above results about $N\phi$ excite our interest in its
analogues, i.e. the systems of $\Omega\pi$, $\Omega\omega$ and
$\Omega\rho$. Similar to the $N\phi$ system, all of them are the
systems where two color singlet clusters have no common flavor
quarks and there are no one-gluon exchange (OGE) interaction between
these two hadrons, thus they are really the special two-hadron
states to examine the quark-quark interactions and further the
interactions between those two hadrons. Here we pay special
attention to the $\Omega\pi$ system, which we think is the most
interesting one. One knows that both $\Omega$ and $\pi$ have the
traits of long life [in order of $10^{-10}$ and $10^{-8}$ seconds],
and there is no coupling channel below the threshold of $\Omega\pi$.
So, if there is really an $\Omega\pi$ bound state, it can not fall
apart into $\Omega$ and $\pi$, and besides, it can not decay via
strong interaction into any other channels, thus its width must be
narrow. Furthermore, the $\Omega\pi^-$ state might be easily
distinguished and detected in the experiments since it has two
negative electronic charges.

It is a general consensus that QCD is the underlying theory of the
strong interaction. However, as the non-perturbative QCD effect is
very important for light quark systems in the low energy region and
it is difficult to be seriously solved, people still need
QCD-inspired models to be a bridge connecting the QCD fundamental
theory and the experimental observables. Among these
phenomenological models, the chiral SU(3) quark model has been quite
successful in reproducing the energies of the baryon ground states,
the binding energy of the deuteron, the nucleon-nucleon ($NN$)
scattering phase shifts, and the hyperon-nucleon ($YN$) cross
sections \cite{zyzhang97}. In this model, the quark-quark
interaction contains confinement, one-gluon exchange (OGE) and boson
exchanges stemming from scalar and pseudoscalar nonets, and the
short range quark-quark interaction is provided by OGE and quark
exchange effects.

Actually it is still a controversial problem in the low-energy
hadron physics whether gluon or Goldstone boson is the proper
effective degree of freedom besides the constituent quark. Glozman
and Riska proposed that the Goldstone boson is the only other proper
effective degree of freedom \cite{glozman96,glozman00}. But Isgur
gave a critique of the boson exchange model and insisted that the
OGE governs the baryon structure \cite{isgur001,isgur002}. Anyway,
it is still an open problem in the low-energy hadron physics whether
OGE or vector-meson exchange is the right mechanism  for describing
the short-range quark-quark interaction, or both of them are
important. Thus the chiral SU(3) quark model has been extended to
include the coupling of the quark and vector chiral fields. The OGE
that plays an important role in the short-range quark-quark
interaction in the original chiral SU(3) is now nearly replaced by
the vector meson exchanges. This model, named the extended chiral
SU(3) quark model, has also been successful in reproducing the
energies of the baryon ground states, the binding energy of the
deuteron, and the nucleon-nucleon scattering phase shifts
\cite{lrdai03}.

Recently, we have extended both the chiral SU(3) quark model and the
extended chiral SU(3) quark model from the study of baryon-baryon
scattering processes to the baryon-meson systems by solving a
resonating group method (RGM) equation
\cite{fhuang04nkdk,fhuang05lksk,fhuang05dklksk,fhuang04kn,fhuang05kne}.
We found that some results are similar to those given by the chiral
unitary approach study, such as that both the $\Delta K$ system with
isospin $I=1$ and the $\Sigma K$ system with $I=1/2$ have quite
strong attractions \cite{fhuang04nkdk,fhuang05lksk,fhuang05dklksk}.
In the study of the $KN$ scattering
\cite{fhuang04nkdk,fhuang04kn,fhuang05kne}, we get a considerable
improvement not only on the signs but also on the magnitudes of the
theoretical phase shifts comparing with other's previous work. We
also studied the phase shifts of $\pi K$ \cite{fhuang05kp}, and got
reasonable fit with the experiments in the low energy region. All
these achievements encourage us to investigate more baryon-meson
systems by using these two models.

In this paper, we dynamically study the interaction of the
$\Omega\pi$ state with $I=1$ and $J^{p}=3/2^-$ in both the chiral
SU(3) quark model and the extended chiral SU(3) quark model by
resolving the RGM equation. The model parameters are taken from our
previous works \cite{fhuang05lksk,fhuang05dklksk,fhuang06nphi},
which gave a satisfactory description of the energies of the baryon
ground states, the binding energy of the deuteron, the $NN$
scattering phase shifts and the hyperon-nucleon ($YN$) cross
sections. The results show that in the extended chiral SU(3) quark
model the $\Omega\pi$ state can be weakly bound as a molecular
state. Experimentally to search this state in the heavy ion
collisions or $e^+e^-$ collisions is worth trying in future.

In Sec. II, we give a brief introduction of the framework of these
two chiral quark models. In Sec. III, the results for the
$\Omega\pi$ state and some discussions are presented. Finally, the
summary is given in Sec. IV.

\section{Formulation}

The chiral SU(3) quark model and the extended chiral SU(3) quark
model has been widely described in the literature
\cite{fhuang04kn,fhuang04nkdk,fhuang05lksk,fhuang05kne,fhuang05dklksk},
and we refer the reader to those works for details. Here we just
give the salient features of these two models.

In these two models, the total Hamiltonian of baryon-meson systems
can be written as
\begin{equation}
H=\sum_{i=1}^{5}T_{i}-T_{G}+\sum_{i<j=1}^{4}V_{ij}+\sum_{i=1}^{4}V_{i\bar
5},
\end{equation}
where $T_G$ is the kinetic energy operator for the center-of-mass
motion, and $V_{ij}$ and $V_{i\bar 5}$ represent the quark-quark
and quark-antiquark interactions, respectively,
\begin{equation}
V_{ij}= V^{OGE}_{ij} + V^{conf}_{ij} + V^{ch}_{ij},
\end{equation}
where $V_{ij}^{OGE}$ is the OGE interaction, and $V_{ij}^{conf}$
is the confinement potential. $V^{ch}_{ij}$ represents the chiral
fields induced effective quark-quark potential. In the chiral
SU(3) quark model, $V^{ch}_{ij}$ includes the scalar boson
exchanges and the pseudoscalar boson exchanges,
\begin{eqnarray}
V^{ch}_{ij} = \sum_{a=0}^8 V_{\sigma_a}({\bm r}_{ij})+\sum_{a=0}^8
V_{\pi_a}({\bm r}_{ij}),
\end{eqnarray}
and in the extended chiral SU(3) quark model, the vector boson
exchanges are also included,
\begin{eqnarray}
V^{ch}_{ij} = \sum_{a=0}^8 V_{\sigma_a}({\bm r}_{ij})+\sum_{a=0}^8
V_{\pi_a}({\bm r}_{ij})+\sum_{a=0}^8 V_{\rho_a}({\bm r}_{ij}).
\end{eqnarray}
Here $\sigma_{0},...,\sigma_{8}$ are the scalar nonet fields,
$\pi_{0},..,\pi_{8}$ the pseudoscalar nonet fields, and
$\rho_{0},..,\rho_{8}$ the vector nonet fields. The expressions of
these potentials can be found in the literature
\cite{fhuang04kn,fhuang04nkdk,fhuang05lksk,fhuang05kne,fhuang05dklksk}.

$V_{i \bar 5}$ in Eq. (1) includes two parts: direct interaction
and annihilation parts,
\begin{equation}
V_{i\bar 5}=V^{dir}_{i\bar 5}+V^{ann}_{i\bar 5},
\end{equation}
with
\begin{equation}
V_{i\bar 5}^{dir}=V_{i\bar 5}^{conf}+V_{i\bar 5}^{OGE}+V_{i\bar
5}^{ch},
\end{equation}
and
\begin{eqnarray}
V_{i\bar{5}}^{ch}=\sum_{j}(-1)^{G_j}V_{i5}^{ch,j}.
\end{eqnarray}
Here $(-1)^{G_j}$ represents the G parity of the $j$th meson. The
$q\bar q$ annihilation interactions, $V_{i\bar 5}^{ann}$, are not
included in this work because they are assumed to be negligible,
because it only acts in very short range region, thus it does not
contribute significantly to a molecular state or a scattering
process which is the subject of our present study.

All the model parameters are taken from our previous work
\cite{fhuang05lksk,fhuang05dklksk}, which gave a satisfactory
description of the energies of the baryon ground states, the
binding energy of deuteron, and the $NN$ scattering phase shifts.
Here we briefly give the procedure for the parameter
determination. The three initial input parameters, i.e. the
harmonic-oscillator width parameter $b_u$, the up (down) quark
mass $m_{u(d)}$ and the strange quark mass $m_s$, are taken to be
the usual values: $b_u=0.5$ fm for the chiral SU(3) quark model
and $0.45$ fm for the extended chiral SU(3) quark model,
$m_{u(d)}=313$ MeV, and $m_s=470$ MeV. The coupling constant for
scalar and pseudoscalar chiral field coupling, $g_{ch}$, is fixed
by the relation
\begin{eqnarray}
\frac{g^{2}_{ch}}{4\pi} = \left( \frac{3}{5} \right)^{2}
\frac{g^{2}_{NN\pi}}{4\pi} \frac{m^{2}_{u}}{M^{2}_{N}},
\end{eqnarray}
with the empirical value $g^{2}_{NN\pi}/4\pi=13.67$. The coupling
constant for vector coupling of the vector-meson field is taken to
be $g_{chv}=2.351$, the same as used in the $NN$ case
\cite{lrdai03}. The masses of the mesons are taken to be the
experimental values, except for the $\sigma$ meson. The $m_\sigma$
is adjusted to fit the binding energy of deuteron. The OGE
coupling constants and the strengths of the confinement potential
are fitted by baryon masses and their stability conditions. All
the parameters are tabulated in Table I, where the first set is
for the chiral SU(3) quark model, the second and third sets are
for the extended chiral SU(3) quark model by taking
$f_{chv}/g_{chv}$ as $0$ and $2/3$, respectively. Here $f_{chv}$
is the coupling constant for tensor coupling of the vector meson
fields.

{\small
\begin{table}[htb]
\caption{\label{para} Model parameters. The meson masses and the
cutoff masses: $m_{\sigma'}=980$ MeV, $m_{\kappa}=980$ MeV,
$m_{\epsilon}=980$ MeV, $m_{\pi}=138$ MeV, $m_K=495$ MeV,
$m_{\eta}=549$ MeV, $m_{\eta'}=957$ MeV, $m_{\rho}=770$ MeV,
$m_{K^*}=892$ MeV, $m_{\omega}=782$ MeV, $m_{\phi}=1020$ MeV, and
$\Lambda=1100$ MeV.}
\begin{center}
\begin{tabular}{cccc}
\hline\hline
  & $\chi$-SU(3) QM & \multicolumn{2}{c}{Ex. $\chi$-SU(3) QM}  \\
  &   I   &    II    &    III \\  \cline{3-4}
  &  & $f_{chv}/g_{chv}=0$ & $f_{chv}/g_{chv}=2/3$ \\
\hline
 $b_u$ (fm)  & 0.5 & 0.45 & 0.45 \\
 $m_u$ (MeV) & 313 & 313 & 313 \\
 $m_s$ (MeV) & 470 & 470 & 470 \\
 $g_u^2$     & 0.766 & 0.056 & 0.132 \\
 $g_s^2$     & 0.846 & 0.203 & 0.250 \\
 $g_{ch}$    & 2.621 & 2.621 & 2.621  \\
 $g_{chv}$   &       & 2.351 & 1.973  \\
 $m_\sigma$ (MeV) & 595 & 535 & 547 \\
 $a^c_{uu}$ (MeV/fm$^2$) & 46.6 & 44.5 & 39.1 \\
 $a^c_{us}$ (MeV/fm$^2$) & 58.7 & 79.6 & 69.2 \\
 $a^c_{ss}$ (MeV/fm$^2$) & 99.2 & 163.7 & 142.5 \\
 $a^{c0}_{uu}$ (MeV)  & $-$42.4 & $-$72.3 & $-$62.9 \\
 $a^{c0}_{us}$ (MeV)  & $-$36.2 & $-$87.6 & $-$74.6 \\
 $a^{c0}_{ss}$ (MeV)  & $-$33.8 & $-$108.0 & $-$91.0 \\
\hline\hline
\end{tabular}
\end{center}
\end{table}}

From Table I one can see that for both set II and set III, $g_u^2$
and $g_s^2$ are much smaller than the values of set I. This means
that in the extended chiral SU(3) quark model, the coupling
constants of OGE are greatly reduced when the coupling of quarks
and vector-meson field is considered. Thus the OGE that plays an
important role of the quark-quark short-range interaction in the
original chiral SU(3) quark model is now nearly replaced by the
vector-meson exchange. In other words, the mechanisms of the
quark-quark short-range interactions in these two models are quite
different.

With all parameters determined, the $\Omega\pi$ state can be
dynamically studied in the framework of the RGM, a well established
method for studying the interaction between two composite particles.
The wave function of the $\Omega\pi$ system is of the form
\begin{eqnarray}
\Psi={\cal A}[{\hat \psi}_\Omega(\bm \xi_1,\bm \xi_2) {\hat
\psi}_\pi(\bm \xi_3) \chi({\bm R}_{\Omega\pi})],
\end{eqnarray}
where ${\bm \xi}_1$ and ${\bm \xi}_2$ are the internal coordinates
for the cluster $\Omega$, and ${\bm \xi}_3$ the internal coordinate
for the cluster $\pi$. ${\bm R}_{\Omega\pi}\equiv {\bm
R}_\Omega-{\bm R}_\pi$ is the relative coordinate between the two
clusters, $\Omega$ and $\pi$. The ${\hat \psi}_\Omega$ is the
antisymmetrized internal cluster wave function of $\Omega$, as well
as ${\hat \psi}_\pi$ is the internal wave function of $\pi$, and
$\chi({\bm R}_{\Omega\pi})$ the relative wave function of the two
clusters. The symbol $\cal A$ is the antisymmetrizing operator
defined as
\begin{equation}
{\cal A}\equiv{1-\sum_{i \in {\Omega}}P_{i4}}\equiv{1-3P_{34}}.
\end{equation}
Expanding unknown $\chi({\bm R}_{\Omega\pi})$ by employing
well-defined basis wave functions, such as Gaussian functions, one
can solve the RGM equation for a bound-state problem or a scattering
one to obtain the binding energy or scattering phase shifts for the
two-cluster systems. The details of solving the RGM equation can be
found in Refs. \cite{wildermuth77,kamimura77,oka81}.

\section{Results and discussions}

\begin{figure}[htb]
\epsfig{file=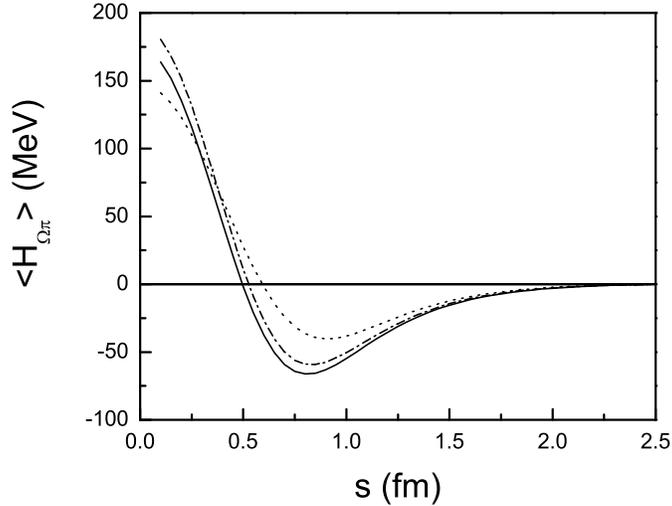,width=9.9cm} \caption{\small The GCM matrix
elements of the Hamiltonian. The dotted, solid and dash-dotted lines
represent the results obtained in model I, II and III,
respectively.}
\end{figure}

As mentioned in the introduction, the $\Omega\pi$ system is a very
special two-hadron state. This is not only because these two
color-singlet clusters have no common flavor quarks and there is no
one-gluon exchange (OGE) interaction between these two hadrons,
which makes this system an ideal place to examine the quark-quark
interactions and further the interactions between those two hadrons,
but also because both $\Omega$ and $\pi$ have long life and narrow
width, and there is no coupling channel below the threshold of
$\Omega\pi$, which means that if there is really an $\Omega\pi$
bound state, its width must be quite narrow. The narrow width and
the two negative electronic charges of the possible $\Omega\pi^-$
bound state can make it easy to distinguish and detect this system
in the experiment. Theoretically, a dynamical investigation of this
interesting light quark system in the framework of the chiral quark
model including the coupling between the quark and chiral fields
seems to be necessary and essential.

Here we study the $\Omega\pi$ state in our chiral quark models by
treating $\Omega$ and $\pi$ as two clusters and solving the
corresponding RGM equation. Fig. 1 shows the diagonal matrix
elements of the Hamiltonian for the $\Omega\pi$ system in the
generator coordinate method (GCM) \cite{wildermuth77} calculation,
which can be regarded as the effective Hamiltonian of two clusters
$\Omega$ and $\pi$ qualitatively. In Fig. 1, $\langle
H_{\Omega\pi}\rangle$ includes the kinetic energy of the relative
motion and the effective potential between $\Omega$ and $\pi$, and
$s$ denotes the generator coordinate which can qualitatively
describe the distance between the two clusters. From Fig. 1, one
sees that the $\Omega\pi$ interaction, dominantly provided by the
$\sigma$ field coupling, is attractive in the medium range. To study
if such an attraction can make for a bound state of the $\Omega\pi$
system, we solve the RGM equation of bound state problem. The
results show that in model I, i.e. the chiral SU(3) quark model, the
$\Omega\pi$ state is unbound. However in model II and III, i.e. the
extended chiral SU(3) quark model with $f_{\rm chv}/g_{\rm chv}=0$
and $f_{\rm chv}/g_{\rm chv}=2/3$, we get weakly bound states of
$\Omega\pi$ with the binding energies of about 2 and 0.5 MeV,
respectively. Actually, as seen in Fig. 1, the $\Omega\pi$
interaction in model II and III are more attractive than those in
model I, thus in model II and III we can get the weakly $\Omega\pi$
bound states while in model I the $\Omega\pi$ state is unbound.
These results are similar to those of the $N\phi$ systems
\cite{fhuang06nphi}.

{\small
\begin{table}[htb] \caption{The binding energy of $\Omega\pi$.}
\begin{tabular*}{120mm}{@{\extracolsep\fill}lccc}
\hline\hline
 & $\chi$-SU(3) QM & \multicolumn{2}{c}{Ex. $\chi$-SU(3) QM }\\
\cline{3-4}
 &  & $f_{\rm chv}/g_{\rm chv}=0$ & $f_{\rm chv}/g_{\rm chv}=2/3$ \\
\hline
$B_{\Omega\pi}$ (MeV) & $-$ & $2$  & $0.5$ \\
\hline\hline
\end{tabular*}
\end{table}}

As the same situation of the $N\phi$ state, in the $\Omega\pi$
system the OGE is not allowed between $\Omega$ and $\pi$ since these
two color-singlet clusters have no common flavor quarks, and the
$\sigma$ exchange dominantly provides the $\Omega\pi$ attractive
interaction in the chiral SU(3) quark model. At the same time, in
the extended chiral SU(3) quark model there is no contribution from
$\rho$, $\omega$ and $\phi$ exchanges, and the attraction in this
special system also dominantly comes from $\sigma$ exchange. In our
calculation, the model parameters are fitted by the $NN$ scattering
phase shifts, and the mass of $\sigma$ is adjusted by fitting the
deuteron's binding energy, thus the value of $m_{\sigma}$ is
somewhat different for the cases I, II and III. In the cases II and
III the masses of the $\sigma$ meson are smaller than that in case
I, which means that in cases II and III the $\Omega\pi$ states get
more attractions than that in case I, thus much more binding
energies of $\Omega\pi$ are obtained in the extended chiral SU(3)
quark model than in the chiral SU(3) quark model.

We also study the structures of $\Omega\omega$ and $\Omega\rho$ in
the chiral quark models by using the same method. Our results show
that in the extended chiral SU(3) quark model with $f_{\rm
chv}/g_{\rm chv}=0$, there is a weakly bound state of $\Omega\rho$
with 0.9 and 2.6 MeV binding energy for $J^P=3/2^-$ and $J^P=5/2^-$,
respectively. The same results can be obtained in the $\Omega\omega$
system. That is, there are also weakly $\Omega\omega$ and
$\Omega\rho$ bound states with $J^P=3/2^-$ and $J^P=5/2^-$ in the
extended chiral SU(3) quark model. But in these two cases the
channel-coupling effects may be important and un-negligible since
the threshold of $\Xi^*K^*$ is very near to those of $\Omega\rho$
and $\Omega\omega$ and the $\Xi^*K$ and $\Xi K^*$ channels also lie
below the $\Omega\rho$ and $\Omega\omega$ thresholds. Hence
$\Omega\rho$ and $\Omega\omega$ might decay via strong interaction
into other channels and consequently, their widths will be
considerably broad and it might be hard to measure these states
experimentally.

Here we'd like to point out that all the results of $\Omega\pi$,
$\Omega\omega$ and $\Omega\rho$ states are quite similar to that of
$N\phi$ \cite{fhuang06nphi}. This is because all these states have
the similar configurations, i.e. they are composed of two
color-singlet hadrons with no common flavor quarks and no OGE, and
their attractions are mostly provided by the $\sigma$ exchange.
Presently the coupling strength of quark and $\sigma$-field is still
an open problem, and people usually use different values in
different cases in order to fit the observables. Experimentally, the
measurement of the $\Omega\pi$ state to examine whether it is bound
or not would be very important for getting more knowledge of the
coupling between quark and $\sigma$ chiral field. We strongly call
for experimentalists to search for this interesting $\Omega\pi$
state in the $e^+e^-$ collisions or heavy ion collisions.

\section{Summary}

In this work, we dynamically studied the $\Omega\pi$ state in the
chiral SU(3) quark model as well as in the extended chiral SU(3)
quark model by solving the RGM equation. All the model parameters
are taken from our previous work, which can give a satisfactory
description of the energies of the baryon ground states, the binding
energy of the deuteron, and the $NN$ scattering phase shifts. The
calculated results show that the $\Omega\pi$ state has an attractive
interaction, which is dominantly provided by the $\sigma$ exchange.
In the extended chiral SU(3) quark model, such an attractive
interaction can make for an $\Omega\pi$ quasi-bound state with
several MeV binding energy. Experimentally whether there is an
$\Omega\pi$ quasi-bound state or resonance state can help us to test
the strength of the coupling between the quark and $\sigma$ chiral
field.

\begin{acknowledgements}
This work was supported in part by the National Natural Science
Foundation of China, Grant No. 10475087.
\end{acknowledgements}


\begin{thebibliography}{99}
\bibitem{brodsky90}S.J. Brodsky, I.A. Schmidt, and G.F.de Teramond, Phys. Rev. Lett. {\bf 64}, 1011 (1990).
\bibitem{hygao01}H. Gao, T.-S. H. Lee, and V. Marinov, Phys. Rev. C {\bf 63}, 022201(R) (2001).
\bibitem{fhuang06nphi}F. Huang, Z.Y. Zhang, and Y.W. Yu, Phys. Rev. C {\bf 73}, 025207 (2006).
\bibitem{zyzhang97}Z.Y. Zhang, Y.W. Yu, P.N. Shen, L.R. Dai, A. Faessler, and U. Straub, Nucl. Phys. A {\bf 625}, 59 (1997).
\bibitem{glozman96}L.Ya. Glozman and D.O. Riska, Phys. Rept. {\bf 268}, 263 (1996).
\bibitem{glozman00}L.Ya. Glozman, Nucl. Phys. A {\bf 663}, 103c (2000).
\bibitem{isgur001}N. Isgur, Phys. Rev. D {\bf 61}, 118501 (2000).
\bibitem{isgur002}N. Isgur, Phys. Rev. D {\bf 62}, 054026 (2000).
\bibitem{lrdai03}L.R. Dai, Z.Y. Zhang, Y.W. Yu, and P. Wang, Nucl. Phys. A {\bf 727}, 321 (2003).
\bibitem{fhuang05kne}F. Huang and Z.Y. Zhang, Phys. Rev. C {\bf 72}, 024003 (2005).
\bibitem{fhuang04kn}F. Huang, Z.Y. Zhang, and Y.W. Yu, Phys. Rev. C {\bf 70}, 044004 (2004).
\bibitem{fhuang04nkdk}F. Huang and Z.Y. Zhang, Phys. Rev. C {\bf 70}, 064004 (2004).
\bibitem{fhuang05lksk}F. Huang, D. Zhang, Z.Y. Zhang, and Y.W. Yu, Phys. Rev. C {\bf 71}, 064001 (2005).
\bibitem{fhuang05dklksk}F. Huang and Z.Y. Zhang, Phys. Rev. C {\bf 72}, 068201 (2005).
%\bibitem{fhuang05np}F. Huang, Z.Y. Zhang, and Y.W. Yu, High Energy Phys. Nucl. Phys. {\bf 29} (2005) 948.
\bibitem{fhuang05kp}F. Huang, Z.Y. Zhang, and Y.W. Yu, Commun. Theor. Phys. {\bf 44}, 665 (2005).
\bibitem{wildermuth77}K. Wildermuth and Y.C. Tang, {\it A Unified Theory of the Nucleus}, Vieweg, Braunschweig (1977).
\bibitem{kamimura77}M. Kamimura, Suppl. Prog. Theor. Phys. {\bf 62}, 236 (1977).
\bibitem{oka81}M. Oka and K. Yazaki, Prog. Theor. Phys. {\bf 66}, 556 (1981).
\end{thebibliography}
\end{document}